\documentclass[aps,prb,twocolumn,groupedaddress,amsmath,amssymb,showpacs]{revtex4-1}
\usepackage{bm}
\usepackage{graphicx}
\newcommand{\com}[2]{\ensuremath{[#1,\;#2]}}
\newcommand{\acom}[2]{\ensuremath{\{#1,\;#2\}}}
\newcommand{\abs}[1]{\ensuremath{|#1|}}
\newcommand{\brac}[1]{\ensuremath{\langle\!\langle #1|}}
\newcommand{\ketc}[1]{\ensuremath{|#1\rangle\!\rangle}}

\newcommand{\ketbrac}[2]{\ensuremath{\ketc{#1}\!\brac{#2}}}

\newcommand{\bra}[1]{\ensuremath{\langle #1|}}
\newcommand{\ket}[1]{\ensuremath{|#1\rangle}}
\newcommand{\ketbra}[2]{\ensuremath{\ket{#1}\!\bra{#2}}}
\newcommand{\tr}{\ensuremath{\operatorname{tr}}}
\newcommand{\ev}[1]{\ensuremath{\langle #1 \rangle}}

\renewcommand{\Im}{\ensuremath{\operatorname{Im}}}

\begin{document}
	\title{Coherent quantum ratchets driven by tunnel oscillations: Fluctuations and correlations}
	\author{Robert Hussein}
	\email[]{robert.hussein@csic.es}
	\author{Sigmund Kohler}
	\affiliation{Instituto de Ciencia de Materiales de Madrid, CSIC, Cantoblanco, E-28049 Madrid, Spain}
	\date{\today}
	
	\begin{abstract}
		We study two capacitively coupled double quantum dots focusing on the
		regime in which one double dot is strongly biased, while no voltage is
		applied to the other.  Then the latter experiences an effective
		driving force which induces a ratchet current, i.e., a dc current in
		the absence of a bias voltage.  Its current noise is investigated with
		a quantum master equation in terms of the full-counting statistics.
		This reveals, that whenever the ratchet current is large, it also
		exhibits some features of a Poissonian process.  By eliminating the
		drive circuit, we obtain a reduced master equation which provides analytical
		results for the Fano factor.
	\end{abstract}
	
	\maketitle
	
	\section{Introduction}
	The realization of double or triple quantum dots in a linear
	arrangement \cite{GaudreauPRL2006a, SchroerPRB2007a, TaubertPRL2008a,
	GrangerPRB2010a} or in a ring configuration \cite{GustavssonPRL2007a,
	RoggePRB2008a} enables transport experiments in which electrons flow
	through delocalized orbitals.  This delocalization is visible in
	modified quadruple points of the charging diagram.
	\cite{GaudreauPRL2006a, SchroerPRB2007a, TaubertPRL2008a, GrangerPRB2010a}  
	Recently, also the capacitive coupling of two double quantum dots has
	been achieved. \cite{KhrapaiPRL2006a, PeterssonPRL2009a, ShinkaiPRL2009a} If
	each double quantum dot is occupied by one electron, the setup
	represents two interacting charge qubits.\cite{PeterssonPRL2009a,
	ShinkaiPRL2009a} Upon opening one double dot, a current flows and may be
	used for readout of the other double dot which still forms a charge
	qubit.  \cite{GiladPRL2006a, JiaoPRB2007a, AshhabPS2009a, KreisbeckPRB2010a}
	
	Opening both double dots enables experiments with two interacting
	mesoscopic currents.  For example, predominantly coherently
	transported electrons perform tunnel oscillations which act on the
	other double dot as an effective ac force.  This may induce a ratchet
	or pump effect, i.e., cause a net current in the absence of any bias
	voltage.\cite{StarkEPL2010a}  A similar ratchet effect has been realized
	by coupling a double dot to a quantum point
	contact.\cite{KhrapaiPRL2006a} This effect is closely related to Coulomb
	drag \cite{MortensenPRL2001a, SanchezPRL2010a} and to using a double quantum dot as noise
	detector.\cite{AguadoPRL2000a, OnacPRL2006a, GustavssonPRL2007a}
	In turn, interacting channels may block each
	other.\cite{GattobigioPRB2002a, BartholdPRL2006a, SanchezPRB2008a}
	
	A common characterization of the current fluctuations in a mesoscopic
	conductor is the full-counting statistics, \cite{LesovikPRL1994a,
	BagretsPRB2003a}  whose cornerstone is a counting variable for the lead
	electrons.  In this way, one obtains a cumulant-generating function
	for the transported charge.  Of particular interest is its variance,
	because it relates to the zero-frequency limit of the current--current
	correlation function.\cite{MacDonaldRPP1949a} Moreover, it indicates
	whether transport is sub or super Poissonian,\cite{BlanterPR2000a} even
	though a more faithful criterion is the $g^{(2)}$
	function.\cite{EmaryPRB2012a} Generally, some further cumulants are
	specific to the system while beyond a certain order, cumulants exhibit
	universal features.\cite{FlindtPNASUSA2009a}
	
	In this work, we explore the noise properties of the ratchet mechanism
	proposed in Ref.~\onlinecite{StarkEPL2010a} for capacitively coupled
	double quantum dots focusing on the full-counting statistics.
	Besides a numerical study with a master equation for the full
	ratchet--drive setup, we derive in the spirit of
	Ref.~\onlinecite{EmaryPRA2008a} an effective master equation for the
	ratchet under the influence of the drive circuit.  This provides an
	analytical expression for the cumulant generating function.  This
	approach is beyond a more heuristic elimination of the drive
	circuit\cite{StarkEPL2010a} and beyond a golden-rule
	calculation,\cite{AguadoPRL2000a} because it includes effects stemming
	from delocalization and from the broadening of the ratchet levels.
	Therefore, it holds also for small ratchet detuning.
	
	\begin{figure}[b]
		\includegraphics{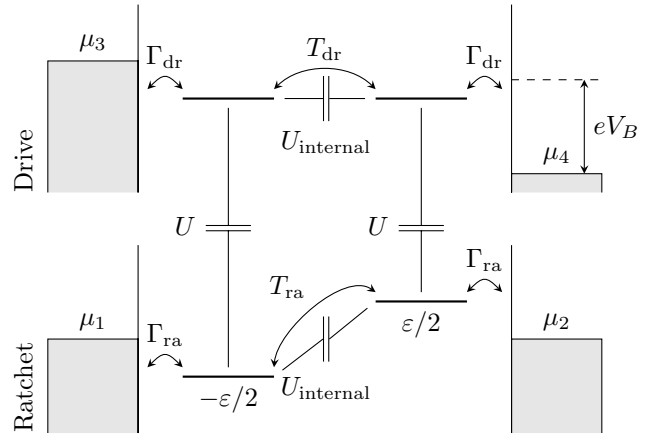}		\caption{\label{fig.:DQS_scheme}
		Quantum ratchet (lower circuit, unbiased) capacitively
		coupled to a drive circuit (top) biased by a voltage $V_B$.  Each
		circuit is modeled as two-level system with tunnel couplings $T_\textrm{ra}$ and
		$T_\textrm{dr}$, respectively.  The ratchet possess a detuning
		$\varepsilon$, while the drive circuit is undetuned.  The dot--lead
		tunnel rates are $\Gamma_{\textrm{ra}}$ and $\Gamma_{\textrm{dr}}$,
		while $\mu_\alpha$, $\alpha=1,\ldots, 4$, denotes the chemical potentials of
		the leads.
		}
	\end{figure}
	In Sec.~\ref{sec.:model}, we introduce our model Hamiltonian for the setup
	sketched in Fig.~\ref{fig.:DQS_scheme} and, moreover, introduce a
	quantum master equation for the full system.
	Section~\ref{sec.:eff_masterequation} is devoted to the elimination of
	the drive circuit which provides our analytical results. In
	Sec.~\ref{sec.:transport_characterization}, we present our numerical
	results for the higher-order cumulants and test the quality of our
	approximations.
	
	\section{\label{sec.:model}Model and method}
	\subsection{Hamiltonian}
	The four quantum dots and the leads (see Fig.~\ref{fig.:DQS_scheme})
	are described by the Hamiltonian
	$\hat H = \hat H_S + \hat H_B + \hat H_V$,
	where 
	\begin{align}
		\begin{split}
\hat H_S ={}
& \sum_{\alpha}\nu_\alpha \hat d_\alpha^\dag \hat d_\alpha
	-\big(T_\textrm{ra} \hat d_1^\dag\hat d_2 + T_\textrm{ra}^* \hat d_2^\dag\hat d_1\big)\\
& -\big(T_\textrm{dr} \hat d_3^\dag\hat d_4 + T_\textrm{dr}^* \hat d_4^\dag\hat d_3\big)
	+\sum_{\alpha<\beta} 
	U_{\alpha\beta}\hat n_\alpha \hat n_\beta
		\end{split}
	\end{align}
	models the quantum dots with the electron creation and
	annihilation operators $\hat d_\alpha^\dagger$ and $\hat d_\alpha$.  The ratchet
	circuit ($\alpha=1,2$) has inter-dot tunneling $T_\textrm{ra}$ and detuning
	$\varepsilon$, such that the level splitting becomes
	\begin{align}
\delta = \sqrt{\varepsilon^2+4 \abs{T_\textrm{ra}}^2}.
	\end{align}
	The levels of the drive circuit ($\alpha=3,4$) are not detuned and
	possess a tunnel matrix element $T_\textrm{dr}$.  The setup is assumed
	symmetric, such that inter-channel Coulomb repulsion reads
	$U\equiv U_{13} = U_{24}$, while the internal repulsions $U_{12}$ and
	$U_{34}$ are assumed so large, that each channel can be occupied with
	at most one electron.  The inter-channel coupling $U$ by contrast, is
	relatively weak but nevertheless is the relevant interaction for
	inducing a ratchet current.  \cite{StarkEPL2010a}
	We do not take into account more indirect interaction mediated by
	phonons \cite{GrangerNatureP2012a} or by a qubit.\cite{NietnerPRB2012a}
	
	Each dot $\alpha$ is coupled to a lead with chemical potential
	$\mu_\alpha$, where $\mu_1=\mu_2$, while $\mu_3\gg\mu_4$, such that
	all levels of the drive circuit lie within the voltage window.
	The lead Hamiltonian and the dot-lead couplings read
	\begin{align}
\hat H_B 
=&\sum_{k\alpha} \varepsilon_{k\alpha} \hat c_{k\alpha}^\dag \hat c_{k\alpha}, \\
\hat H_V
=&\sum_{k\alpha} \big(
	V_{k\alpha} \hat c_{k\alpha}^\dag \hat d_\alpha 
	+V_{k\alpha}^* \hat d_\alpha^\dag \hat c_{k\alpha} \big),
	\end{align}
	respectively, where $\hat c^\dag_{k\alpha} $and $\hat c_{k\alpha}$ are
	the fermionic operators and $\varepsilon_{k\alpha}$ is the corresponding
	single-particle energy.  The system--bath interaction $\hat H_V$
	is determined by the effective tunnel rates $\Gamma_\alpha(\varepsilon) =
	2\pi\sum_{k}\abs{V_{k\alpha}}^2\delta(\varepsilon-\varepsilon_{k\alpha})
	\equiv\Gamma_\alpha$, which within the wide-band limit are assumed
	energy-independent. Throughout this work we use units in which
	$\hbar=1$.

	\subsection{Cumulant generating function and master equation}
	
	We are interested in the low-frequency properties of the transport
	process, which can be characterized by the distribution of the number of
	transported electrons at large times or, equivalently, by the
	corresponding cumulants.  We thus introduce for each lead $\alpha$ a
	counting variable $\chi_\alpha$ such that we obtain the moment generating
	function $\ev{e^{i\bm\chi (\hat{\bm  N}-\bm N_0)}}_t \equiv \exp[G(\bm{\chi},t)]$,
	where $\bm\chi = (\chi_1, \chi_2, \chi_3, \chi_4)$, while $\bm N =
	(N_1,N_2, N_3, N_4)$ is the electron number in each lead in vector
	notation with the initial value $\bm N_0=\ev{\hat{\bm  N}}_{t=0}$.
	Obviously, the Taylor coefficients of $\exp[G(\bm{\chi},t)]$ are
	the moments of the lead electron distributions.  This allows the
	definition of occupation
	cumulants as Taylor coefficients of $\ln\ev{e^{i\bm\chi (\hat{\bm  N}-\bm N_0)}}_t$.
	Eventually they grow linearly in time.\cite{BagretsPRB2003a}  Thus, the
	information about the stationary limit is contained in the time
	derivative in the long-time limit, so that the (particle) current
	cumulants of leads 2 and 4 can be written as
	\begin{align}
\kappa_{m,n}
&= \frac{\partial^m}{\partial(i \chi_2)^m} \frac{\partial^n}{\partial(i \chi_4)^n}
   \frac{\partial}{\partial t} G(\bm{\chi},t)\Big|_{\bm{\chi}\to0,
t\to\infty} \label{eq.:DQS_current_cumulant}.
	\end{align}
	Owing to charge conservation, the low-frequency properties of the
	currents in leads 1 and 3 are identical with those of leads 2 and 4,
	respectively.  Thus, it is sufficient to consider only the latter.
	The first-order current cumulants are the 
	ratchet current $I_\textrm{ra} = e_0\kappa_{1,0}$ and the drive current
	$I_\textrm{dr} = e_0\kappa_{0,1}$, where $e_0$ denotes the elementary
	charge.  The second-order derivatives
	correspond to the zero-frequency limit of the current-current
	correlation functions,\cite{MacDonaldRPP1949a} in particular, 
	$S_\textrm{ra} = e_0^2\kappa_{2,0}$.
	Since our focus lies on the ratchet current, we introduce the notation
	$\kappa_{m}\equiv \kappa_{m,0}$.
	
	The cumulant generating function $G(\bm\chi,t)$ can be obtained from a
	Markovian master equation approach by unraveling the reduced density
	operator according to the number of electrons $N_\alpha$ in the leads.
	\cite{BagretsPRB2003a}  Alternatively, one may attribute a counting
	variable to the system--lead tunnel operators \cite{LevitovJMP1996a,
	GogolinPRB2006a, SchallerPRB2009a} or multiply the full density
	operator by $e^{i\bm\chi\cdot\hat{\bm N}}$ before tracing out the
	leads.\cite{KaiserAP2007a}  The resulting augmented density operator
	$\rho(\bm\chi,t)$ relates to the cumulant generating function
	via $\ln[\tr\rho(\bm\chi,t)] = G(\bm\chi,t)$, while its
	limit $\rho(\bm\chi\to\bm 0,t)$ is the usual reduced density operator.
	Moreover, $\rho(\bm{\chi},t)$ obeys the master equation
	\begin{align}
\dot\rho(\bm{\chi},t) 
	&=\mathcal L(\bm{\chi})\rho(\bm{\chi},t)\nonumber\\
	&\equiv \Big[
		\mathcal L_0 +\sum_{\alpha,s=\pm}(e^{s i\chi_\alpha}-1)\mathcal J^s_\alpha
		\Big] \rho(\bm{\chi},t) \label{eq.:DQS_masterequation},
	\end{align}
	where $\mathcal L_0$ is the Liouville operator, in our case the one
	obtained within Bloch-Redfield approximation.  For a short derivation,
	see Appendix~\ref{sec.:Liouvillian}.  The superoperator
	$\mathcal J^{+}_\alpha$ describes tunneling of an electron from dot
	$\alpha$ to the respective lead, while $\mathcal J^{-}_\alpha$
	captures the opposite process.  Thus, the counting variables keep
	track of the electron numbers in the leads despite that the latter are
	traced out.  From the master equation \eqref{eq.:DQS_masterequation},
	one can obtain numerically all cumulants within the recursive scheme
	of Ref.~\onlinecite{FlindtPRL2008a}.  Before discussing these results,
	we aim at further analytical progress.

	\section{\label{sec.:eff_masterequation}Elimination of the drive circuit}
	
	In order to reduce the number of degrees of freedom, such that
	an analytically solvable master equation emerges, we eliminate the
	drive circuit along the lines of Ref.~\onlinecite{EmaryPRA2008a}.  We
	start by separating the master equation for $\rho(\bm\chi,t)$,
	Eq.~\eqref{eq.:DQS_masterequation}, into contributions for the
	ratchet, the drive circuit, and their mutual interaction,
	\begin{align}
\dot\rho(\chi_2,t) 
	&= \big[
		\mathcal L_{\textrm{ra}}(\chi_2) + \mathcal L_{\textrm{dr}} + U \mathcal L_{\textrm{ra--dr}}
		\big]\rho(\chi_2,t) \label{eq.:pert_masterequation}.
	\end{align}
	Since we focus on the ratchet current, we keep only the counting variable
	$\chi_2$ for the right lead of the ratchet circuit.  The interaction
	Liouvillian
	\begin{align}
\mathcal L_{\textrm{ra--dr}}\rho
	= -\frac{i}{2}\com{\Delta\hat n_{\textrm{dr}}\Delta\hat n_{\textrm{ra}}}{\rho},
	\end{align}
	is governed by the occupation imbalances $\Delta \hat n_\textrm{ra} = \hat
	n_2 - \hat n_1$ and $\Delta \hat n_\textrm{dr} = \hat n_4-\hat n_3$,
	which allow one to approximately write the ratchet--drive interaction
	Hamiltonian as \cite{StarkEPL2010a} $U (\hat n_1\hat n_3 + \hat
	n_2\hat n_4)\approx (U/2) \Delta \hat n_\textrm{dr}\Delta \hat
	n_\textrm{ra}$.  Thereby we neglect terms that cause global shifts of
	all dot energies.  They are not relevant here, because for all
	parameters considered below, the onsite energies stay far from the
	Fermi surfaces.
	
	After transforming master equation~\eqref{eq.:pert_masterequation} into
	Laplace space, straightforward algebra \cite{EmaryPRA2008a} provides
	for the ratchet an effective Liouvillian $\mathcal L_{\textrm{eff}}$,
	which follows from the relation
	\begin{align}
[z -\mathcal L_{\textrm{eff}}(\chi_2,z)]^{-1} 
&\equiv \tr_{\textrm{dr}}\{[z -\mathcal L(\chi_2)]^{-1} \rho_{\textrm{dr}}^{\textrm{stat}}\}
\label{eq.:conditional_equation_Leff}
	\end{align}
	and depends on the stationary state $\rho_\textrm{dr}^\textrm{stat}$ of
	the drive circuit.  Taylor expansion up to second order in the
	interaction constant $U$ and subsequent evaluation of the partial
	trace yields
	\begin{align}
\mathcal L_{\textrm{eff}}(\chi_2,z) 
&=	\mathcal L_{\textrm{ra}}(\chi_2)
	+ U \mathcal L_{\textrm{eff}}^{(1)}(z) 
	+ U^2 \mathcal L_{\textrm{eff}}^{(2)}(z) 
	\end{align}
	with the linear and quadratic corrections
	\begin{align}
\label{Leff1}
\mathcal L_{\textrm{eff}}^{(1)}(z)
=& -\frac{i}{2}\ev{\Delta\hat n_{\textrm{dr}}}\com{\Delta\hat n_{\textrm{ra}}}{\bullet } ,
\\
\label{Leff2}
		\begin{split}
\mathcal L_{\textrm{eff}}^{(2)}(z)
=& -\frac{1}{4}\sum_m
	C(z -\lambda_{\textrm{ra}}^{(m)})
\\
&\times\com{\Delta\hat n_{\textrm{ra}}}{\bullet}
	\ketbrac{\phi_{\textrm{ra}}^{(m)}}{\tilde\phi_{\textrm{ra}}^{(m)}}
	\com{\Delta\hat n_{\textrm{ra}}}{\bullet}.
		\end{split}
	\end{align}
	Here we employ the superoperator notation of
	Ref.~\onlinecite{Carmichael2008a} and define $[M,\bullet]\rho
	\equiv [M,\rho]$.  Moreover, we have introduced the spectral decomposition
	of the ratchet Liouvillian, $\sum_m \lambda_{\textrm{ra}}^{(m)}
	\ketbrac{\phi_{\textrm{ra}}^{(m)}}{\tilde\phi_{\textrm{ra}}^{(m)}}$,
	with the eigenvalues $\lambda^{(m)} = 0, -\Gamma_\textrm{ra},
	-\Gamma_\textrm{ra}/2\pm i\delta$, and the left and right eigenvectors
	$\brac{\tilde\phi_\textrm{ra}^{(m)}}$ and
	$\ketc{\phi_{\textrm{ra}}^{(m)}}$, respectively.
	Here a difficulty arises from the fact that the Liouvillian of a
	double quantum dot in the zero-bias limit is defective, i.e., it does
	not possess a complete set of eigenvectors (for details, see
	Appendix~\ref{sec.:zero_bias}).  Then one may proceed either by
	constructing a generalized eigenbasis or by introducing a small
	perturbation that lifts the defectiveness, and finally consider the
	limit of vanishing perturbation.  \cite{MolerSIAMReview1978a}
	Since all levels are assumed to stay far from the Fermi surfaces, the
	impact of the interaction on the jump operators can be neglected
	safely.  Thus, the jump operators $\mathcal{J}_2^\pm$ of the effective
	model coincide with the ones of the full Liouvillian.
	
	The dependence on the drive circuit enters via the Laplace transformed
	auto correlation function of the population imbalance
	\begin{align}
C(t) = \ev{\Delta\tilde n_{\textrm{dr}}(t)\Delta\tilde n_\textrm{dr}(0)}
	- \ev{\Delta\hat n_{\textrm{dr}}}^2
	\end{align}
	evaluated at the eigenvalues of the ratchet Liouvillian,
	$C(z-\lambda_\textrm{ra}^{(m)})$, which fulfills $C^*(z) = C(z^*)$.
	For a derivation, see Appendix~\ref{sec.:occupation_cumulant}.  Below
	we will find that the poles of $C(z)$ are related to the extrema of
	the ratchet current.
	
	The linear contribution to the effective Liouvillian,
	$\mathcal{L}_\textrm{eff}^{(1)}$ merely provides a small additional bias for
	the ratchet circuit, but does not induce any non-equilibrium effect.
	Thus, we omit this term and focus on the impact of
	$\mathcal{L}_\textrm{eff}^{(2)}$.  By a straightforward but tedious
	evaluation of Eq.~\eqref{Leff2}, we obtain in the Fock basis
	$\{\ketbra{0}{0}, \ketbra{1}{1}, \ketbra{2}{2},
	\ketbra{2}{1}, \ketbra{1}{2}\}$ the expression
	\begin{align}
\mathcal{L}_\textrm{eff}^{(2)}(z) = 
			\begin{pmatrix}
		0 & 0 & 0 & 0 & 0\\
		0 & 0 & 0 & 0 & 0\\
		0 & 0 & 0 & 0 & 0\\
		0 & 0 & 0 & A(z) & B^*(z^*)\\
		0 & 0 & 0 & B(z) & A^*(z^*)
			\end{pmatrix} ,
	\end{align}
	where the terms
	\begin{align}
	A(z) = &-\frac{\abs{T}^2}{\delta^2}\big[ 2C(z+\Gamma_\textrm{ra}) - \Gamma_\textrm{ra} C'(z+\Gamma_\textrm{ra})\big]\nonumber\\
		&-\sum_{s=\pm} \frac{(\delta+s\varepsilon)^2}{4\delta^2}C(z+\Gamma_\textrm{ra}/2+si\delta),\\
	B(z) = &\frac{T^2}{\delta^2}\big[ 2C(z+\Gamma_\textrm{ra}) - \Gamma_\textrm{ra} C'(z+\Gamma_\textrm{ra})\nonumber\\
		&-C(z+\Gamma_\textrm{ra}/2+i\delta)
		-C(z+\Gamma_\textrm{ra}/2-i\delta)\big]
	\end{align}
	contain a non-Markovian correction through the dependence on the
	Laplace variable $z$.
	For the resulting effective ratchet Liouvillian,
	$\mathcal{L}_\textrm{ra}(\chi_2) +U^2\mathcal{L}_\textrm{eff}^{(2)}(z)$, the
	cumulant generating function can be obtained by a standard procedure,
	namely by computing the eigenvalue that vanishes for $\chi_2\to0$.\cite{BagretsPRB2003a} 
	This yields a somewhat bulky expression and, thus, 
	we restrict ourselves to the Markovian limit obtained by
	$z\to 0$.  By differentiation with respect to $\chi_2$, we obtain the
	current and the zero-frequency noise as
	\begin{align}
\label{Ianalyt}
I_{\textrm{ra}} 
=&{} \frac{e_02b\varepsilon}{\delta} \Im\big[
	\big( -\Gamma_{\textrm{ra}}/2 +i\delta \big)
	C\big( \Gamma_{\textrm{ra}}/2 +i\delta \big)
\big],
\\
\label{Sanalyt}
S_{\textrm{ra}} 
=&{}\frac{e_0^2b}{\delta^2} \Im\big[
		(-\Gamma_{\textrm{ra}}\varepsilon^2 +i\delta\varepsilon^2  +i\delta^3)
		C( \Gamma_{\textrm{ra}}/2 +i\delta)
\big],
	\end{align}
	where $b = {4\abs{T_\textrm{ra}}^2 U^2}/{\delta
	(4\delta^2+\Gamma_{\textrm{ra}}^2)}$.  These expression, in turn,
	allow us to simplify the cumulant generating function $G(\chi_2,t)$.  
	After evaluating the time derivative and the limit $t\to\infty$
	contained in definition~\eqref{eq.:DQS_current_cumulant} of the
	current-cumulant, we obtain the generating function
	\begin{align}
G_{\textrm{eff}}^{(I)}(\chi_2)
	&=\frac{i(I_{\textrm{ra}}/e_0)\sin(\chi_2) + (S_{\textrm{ra}}/e_0^2)[\cos(\chi_2)-1]}{1 
		+\frac{b\Gamma_{\textrm{ra}}^2}{2\delta U^2}[\cos(\chi_2)-1]} ,
\label{eq.:cfg_Leff}
	\end{align}
	which within the present approximation contains the full information
	about the low-frequency properties of the ratchet current.  The
	presence of the counting variable $\chi_2$ in the denominator,
	however, renders the actual calculation of higher-order cumulants a formidable
	task.  Only in the golden-rule limit, i.e., to lowest order in
	$\Gamma_\textrm{ra}$, the denominator becomes independent of $\chi_2$,
	so that $G_\text{eff}^{(I)}(\chi_2) = i(I_\text{ra}^{(0)}/e_0)\sin(\chi_2) +
	(S_\text{ra}^{(0)}/e_0^2)[\cos(\chi_2)-1]$.  Consequently, we obtain the current cumulants
	\begin{equation}
		\label{kappa-analyt}
		\kappa_n
		= \frac{\partial^n }{\partial (i\chi_2)^n}G_\text{eff}^{(I)}(\chi_2)\Big|_{\chi_2=0}
		= \begin{cases}			I_\text{ra}^{(0)}/e_0 & \text{for odd $n$} \\
			S_\text{ra}^{(0)}/e_0^2 & \text{for even $n$}
		\end{cases}
	\end{equation}
	where the upper index $(0)$ refers to the limit $\Gamma_\textrm{ra}\to0$. 
	It turns out to be a good approximation, unless universal cumulant
	oscillations set in, as we will discuss in Sec.~\ref{sec.:cumulantCharacterisation}.
	
	\begin{figure}
		\includegraphics{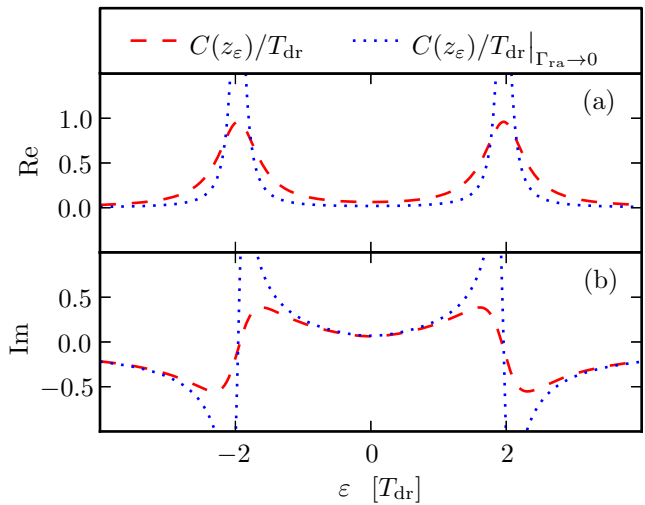}		\caption{\label{fig.:currentCharacterisation}
		(a) Real and (b) imaginary part of the auto correlation function (dashed lines) of the drive population imbalance in
		Laplace representation, $C(z_\varepsilon )$, evaluated at the
		broadened resonance of the ratchet Liouvillian, $z_\varepsilon =
		\Gamma_\textrm{ra}/2 + i({\varepsilon^2 + 4\abs{T_\textrm{ra}}^2})^{1/2}$ 
		with $\Gamma_{\textrm{ra}} = 0.5\,T_\textrm{dr}$ and $T_\textrm{ra} = 0.2\,T_\textrm{dr}$, as function of the
		detuning $\varepsilon$.  The dotted lines correspond to the limit
		$\Gamma_\textrm{ra}\to 0$.
		The dot--lead coupling is $\Gamma_{\textrm{dr}} = T_\textrm{ra}$.
		}
	\end{figure}
	Before testing the
	quality of this approximation and the parameter dependence of the
	results, we close this section by a remark on a formal aspect of
	the perturbation theory.
	Both the current \eqref{Ianalyt} and the zero-frequency
	noise~\eqref{Sanalyt} are proportional to the auto correlation
	function of the population imbalance, $C(z)$, evaluated at the
	broadened level splitting of the ratchet, where the Laplace variable
	reads $z_\varepsilon =
	\Gamma_\textrm{ra}/2+i(\varepsilon^2+4\abs{T_\textrm{ra}}^2)^{1/2}$.  Thus we
	expect the ratchet current to exhibit resonance peaks.  Taking into
	account the broadening distinguishes the present result from that of
	Ref.~\onlinecite{StarkEPL2010a}.  There the ratchet current has been
	computed from the golden-rule rates for noise-induced transitions
	between ratchet eigenstates.  While this treatment accounts properly
	for delocalization effects, it predicts too pronounced resonance
	peaks.  Formally, the golden-rule solution is restored by the
	replacement $C(\Gamma_\text{ra}/2+i\delta)\to C(i\delta)$ in
	Eq.~\eqref{Ianalyt}.  Figure~\ref{fig.:currentCharacterisation}
	visualizes that for ratchet detunings close to resonances,
	the difference between the two approximations may be significant.
	
	\section{\label{sec.:transport_characterization}Characterization of the ratchet current}
	
	\begin{figure}
		\includegraphics{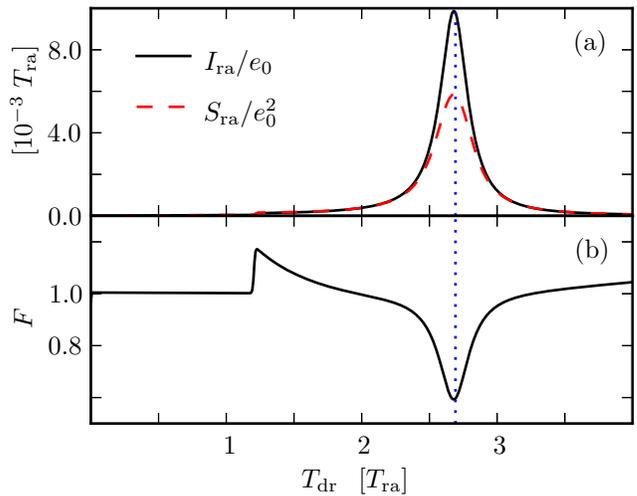}		\caption{\label{fig.:resonantTransport}
		(a) Ratchet current $I_{\textrm{ra}}$, its zero-frequency
		noise $S_{\textrm{ra}}$, and (b) the resulting Fano factor
		$F=S_{\textrm{ra}}/e_0\abs{I_{\textrm{ra}}}$ as function of the
		tunnel matrix element of the drive circuit.  The results are computed
		with the full master equation.  The other parameters are
		$\Gamma_{\textrm{ra}} = \Gamma_{\textrm{dr}} = 0.1\;T_\textrm{ra}$, $U = 0.5\;T_\textrm{ra}$,
		and $\varepsilon=5\;T_\textrm{ra}$.  The vertical dotted line marks the resonance
		condition $4|T_\text{dr}|^2 = \varepsilon^2+4|T_\textrm{ra}|^2$.
		}
	\end{figure}
	\begin{figure*}
		\includegraphics{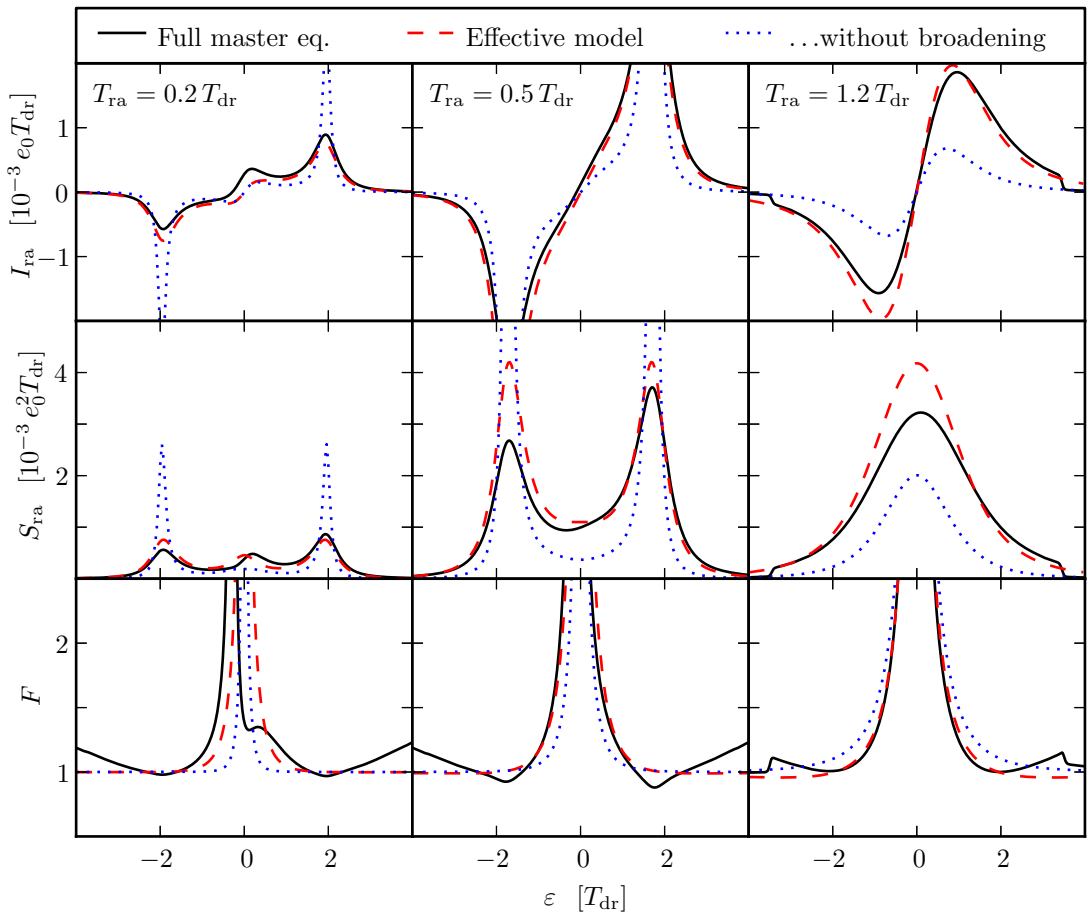}		\caption{\label{fig.:correlationEvolution}
		Ratchet current (upper row), zero-frequency noise (middle),
		and Fano factor (lower row) as function of the ratchet detuning
		$\varepsilon$ for various tunnel couplings $T_\textrm{ra}$.  Results for the full
		master equation (solid lines) are compared to the analytical
		results~\eqref{Ianalyt} and \eqref{Sanalyt} (dashed).  Dotted lines
		mark the golden-rule results which ignore the broadening $\Gamma_\text{ra}$.  The
		dot-lead tunneling rates are $\Gamma_{\textrm{ra}} = 0.5\,T_\textrm{dr}$ and
		$\Gamma_{\textrm{dr}} = 0.2\;T_\textrm{dr}$, while the inter-channel coupling
		reads $U = 0.2\,T_\textrm{dr}$.
		}
	\end{figure*}
	Before starting with the analysis of the ratchet current fluctuations,
	let us compare the present case to that of a ratchet driven by an ac
	field.  There, the current exhibits resonance peaks with large
	current and low zero-frequency noise.\cite{StrassPRL2005a,
	KaiserAP2007a}  For large driving amplitudes, the same behavior is
	visible at multi-photon resonances.
	Figure~\ref{fig.:resonantTransport} shows the corresponding result for
	the present driving by tunnel oscillations.  When the level splitting of the
	ratchet matches the tunnel frequency of the drive circuit, we indeed
	observe the qualitatively same behavior.  Here however, we do not find
	higher-order resonances and, moreover, the Fano factor does not reach the
	extremely small values found in Ref.~\onlinecite{StrassPRL2005a}.  The
	reason for this is that for realistic parameters, the driving via
	Coulomb interaction with the upper circuit is much weaker than direct
	ac driving by, e.g., a high-frequency gate voltage.\cite{Stehlik2012a}
	The kink in the Fano factor stems from a small step in the current and
	can be attributed to an energy difference of many-particle states that
	crosses the Fermi level of the ratchet. \cite{StarkEPL2010a}
	This confirms our picture in which the tunnel oscillations of
	electrons in the drive circuit act like an ac driving with (angular)
	frequency $\Omega = |2T_\text{dr}|$ determined by the tunnel
	splitting.
	
	\subsection{Zero-frequency noise and Fano factor}
	If the tunnel coupling of the ratchet is smaller than that of the
	drive circuit, $T_\text{ra}<T_\text{dr}$, one can adjust the ratchet
	bias $\varepsilon$ such, that the resonance condition $ \varepsilon^2 +
	4\abs{T_\textrm{ra}}^2 =4\abs{T_\textrm{dr}}^2$ is met.  By contrast,
	for $T_\text{ra}>T_\text{dr}$ this is not the case.  In order to first
	sketch the global behavior, we first consider the current, the
	zero-frequency noise, and the resulting Fano factor in dependence of
	the ratchet bias.  We compare numerical results obtained from the full
	master equation with the analytical solution of
	Sec.~\ref{sec.:eff_masterequation}.  Moreover, we also discuss the
	analytical expressions \eqref{Ianalyt} and \eqref{Sanalyt} to lowest
	order in $\Gamma_\text{ra}$, because this restores the golden-rule
	results of Ref.~\onlinecite{StarkEPL2010a}.
	
	Figure~\ref{fig.:correlationEvolution} provides an overview to the
	behavior.  The current which is depicted in the first row, exhibits
	the expected resonance peaks provided that $T_\text{ra}<T_\text{dr}$.
	If the tunnel matrix element is rather small
	($T_\text{ra}=0.2\,T_\text{dr}$), we witness also the small peaks at
	small values of $\varepsilon$, which we predicted within our
	analytical treatment.  Upon increasing the inter-dot tunneling
	$T_\text{ra}$, the current peaks naturally increase as well.  Once
	$T_\text{ra}>T_\text{dr}$, the resonance peaks fade away while the
	structure at $\varepsilon\approx0$ becomes rather pronounced.  In all
	regimes, the analytical result \eqref{Ianalyt} for the current is well
	confirmed.  The main difference is the absence of the slight asymmetry
	with respect to reverting the detuning, $\varepsilon\to -\varepsilon$.
	Nevertheless, the characteristics of the current as function of
	$\varepsilon$ is by and large antisymmetric, which implies a current
	reversal close to zero detuning.
	In order to capture also the lack of perfect antisymmetry, we would
	have to consider the linear perturbation
	$\mathcal{L}_\text{eff}^{(1)}$ which, however, would impede
	concise analytical results.
	For the parameters used in Fig.~\eqref{fig.:correlationEvolution}, the
	golden-rule expression of Ref.~\onlinecite{StarkEPL2010a}, i.e.,
	Eq.~\eqref{Ianalyt} to lowest order in $\Gamma_\text{ra}$,
	reproduces the behavior only qualitatively.  It predicts too sharp
	peaks, because this approximation does not account for the level
	broadening of the ratchet.  The deviation is quite significant in the
	non-resonant case $T_\text{ra}>T_\text{dr}$.
	
	The main features such as the location of the peaks are
	also found for the zero-frequency noise plotted in the middle row.  An
	important difference is found only close to $\varepsilon=0$, where the
	current vanishes for symmetry reasons.  The noise nevertheless remains
	finite and may even have a peak.  This behavior is reflected by the Fano factor
	which stays close to the Poissonian value $F=1$ for detunings far from
	the current reversal point
	$\varepsilon=0$. There the current vanishes, while the noise remains finite,
	such that $F$ diverges.  The reason for this universal behavior can be
	understood from the analytical results for the current and the noise.
	Both $I_\text{ra}$ and $S_\text{ra}$ depend on the drive circuit via
	the drive correlation function $C(z)$ which, thus, cancels in their
	ratio, i.e., in the Fano factor.  On a smaller scale, we observe in
	the Fano factor occasional kinks in less important regions in which
	the current is rather small.  There a small change in the denominator
	of $F = S_\text{ra}/e_0|I_\text{ra}|$ may have a strong effect.

	\begin{figure}
		\includegraphics{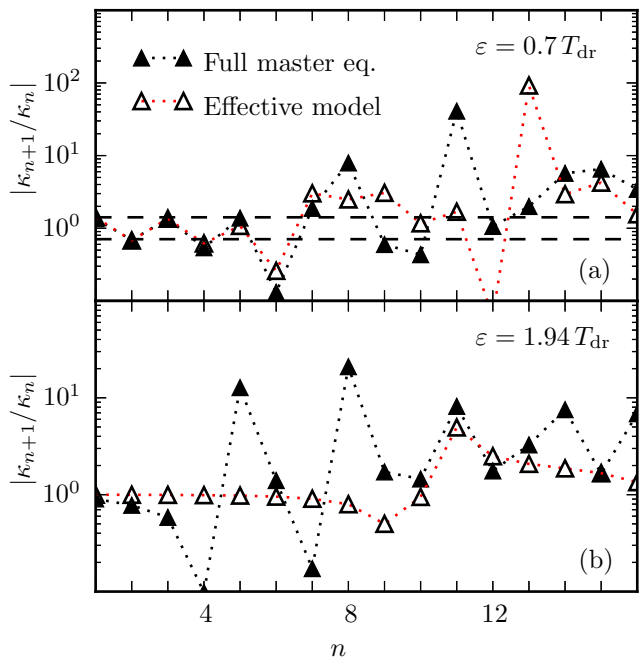}		\caption{\label{fig.:cumulantCharacterisation}
		Cumulant ratio $\abs{\kappa_{n+1}/\kappa_n}$ versus order $n$
		for the parameters used in the second column of
		Fig.~\ref{fig.:correlationEvolution} for two values of the detuning
		$\varepsilon$.  The value $\varepsilon\approx1.94\,T_\text{dr}$
		corresponds to the resonance between ratchet and drive circuit.  The
		horizontal lines in panel (a) mark the analytical result
		\eqref{kappa-analyt} valid to lowest order in
		$\Gamma_\text{ra}$, i.e., $F$ and $1/F$.  The dotted lines serve
		as guide to the eye.
		}
	\end{figure}
	\begin{figure}
		\includegraphics{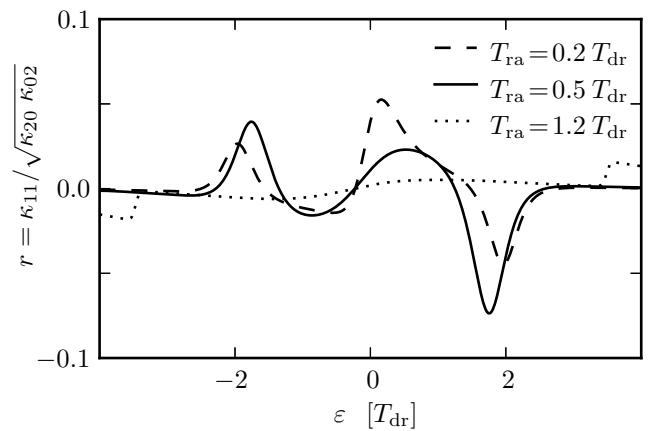}		\caption{\label{fig.:correlationCoefficient}
		Correlation coefficient $r$ versus detuning $\varepsilon$ for
		the parameters used in Fig.~\ref{fig.:correlationEvolution}.
		}
	\end{figure}
	
	\subsection{Higher-order cumulants}
	\label{sec.:cumulantCharacterisation}
	For a refined study of the current noise, we investigate also the
	cumulants of higher order, where we express the results in terms of
	the ratio between subsequent cumulants, $\abs{\kappa_{n+1}/\kappa_n}$.
	For this quantity, the limit of small $\Gamma_\text{ra}$ is rather
	interesting, because our analytical result \eqref{kappa-analyt}
	implies that the cumulant ratio alternates between the Fano factor and
	its reciprocal.  Such behavior is characteristic for a bi-directional Poisson
	process.\cite{LevitovPRB2004a}  The usual Poisson process with
	$|\kappa_{n+1}/\kappa_n|=\text{const}$
	emerges as special case if the backward current is negligible.
	The higher-order cumulants of the effective Liouvillian for larger
	values of $\Gamma_\text{ra}$ can be evaluated from the generating
	function \eqref{eq.:cfg_Leff}, but the expressions become
	rather bulky, so that one has to resort to a numerical evaluation.
	
	Figure~\ref{fig.:cumulantCharacterisation} shows a comparison of these
	two approximations together with the result of the full master equation.  For a
	small ratchet detuning below the resonance [panel(a)], we find that
	all three solutions agree quite well and that the first few cumulants
	exhibit the predicted alternation between the values $F$ and $1/F$.
	For higher orders, the generic universal cumulant oscillations set
	in,\cite{FlindtPNASUSA2009a} which obviously is beyond our analytical
	approach.  At the resonance, the universal oscillations start even
	already at lower order and Eq.~\eqref{kappa-analyt} no longer
	holds.  This is in agreement with our earlier observation that the
	broadening of the ratchet levels plays a significant role for resonant
	driving.

	\subsection{\label{sec.:cross_correlations} Cross correlations}
	Let us finally consider the cross correlations between the drive
	current and the ratchet current.  They can be characterized by the
	cumulant $\kappa_{1,1}$, which is equivalent to the covariance of the
	transported charge in the two circuits.  As a dimensionless measure,
	we introduce the correlation coefficient $r \equiv \kappa_{11}
	/\sqrt{\kappa_{20}\;\kappa_{02}}$, which is bounded by $-1\leq r \leq
	1$.  The results depicted in Fig.~\ref{fig.:correlationCoefficient}
	demonstrate that the correlation between the two currents is rather
	low.  While it can be up to $\abs{r}\sim 0.1$ at the resonances, it is
	hardly noticeable in the non-resonant case $T_\textrm{ra}>T_\text{dr}$.
	
	\section{Conclusions}
	
	A double quantum dot with detuned energy levels may act as quantum
	ratchet or quantum pump when driven out of equilibrium.  An external
	force with zero net bias acting locally upon such system can induce a
	dc current.  Here we investigated a ratchet with a particular driving,
	namely one that stems from the capacitive coupling to a further double quantum
	dot which, however, is biased.  Electrons flowing through the drive
	circuit perform tunnel oscillations which indeed induce phenomena
	similar to those induced by deterministic ac driving.  In this work we
	mainly focused on the fluctuations of the emerging ratchet current.
	Besides a numerical solution with a master equation for all four
	quantum dots, we derived an effective ratchet Liouvillian by
	eliminating the drive circuit.  In this way, we obtained analytical
	results even for higher-order cumulants, which agree well with those
	of the full master equation provided that the tunnel splitting of the
	drive circuit is larger than that of the ratchet.
	
	As a common feature of driving by tunnel oscillations and driving by
	an ac field, we found resonance peaks at which the current assumes a
	maximum, while the relative noise characterized by the Fano factor is
	minimal.  However, clearly sub-Poissonian noise is only found for
	large detuning of the ratchet levels.  This noise reduction should be
	measurable, even though it is not as pronounced as in the case of ac
	driving, mainly because it requires large driving amplitudes
	which cannot be achieved by capacitive coupling.
	For less detuned ratchet levels, the Fano factor is typically of the order one, unless
	the detuning is so small that its orbitals are fully delocalized.
	Then the lack of sufficiently strong asymmetry keeps the
	current at a low value, while the zero-frequency noise stays finite.
	Thus, the Fano factor being the ratio of these two quantities assumes
	very large values.  This generic behavior of the Fano factor is
	explained by our analytical results which reveal that both the current
	and the zero-frequency noise are proportional to the correlation function of
	the drive circuit.  Thus the Fano factor depends only on the shape of
	the ratchet eigenfunctions, while the correlation function cancels.
	
	The higher-order cumulants tend to alternate between two values.  This
	indicates a bi-directional Poisson process and implies that a
	backward current becomes relevant.  With increasing order, however,
	universal oscillations with ever larger amplitude dominate.  The onset
	of the universal oscillations marks the point at which our
	analytically obtained higher-order cumulants significantly deviate
	from those for the full master equation.  Nevertheless, the physically
	relevant cumulants of lower order are well within our analytical
	treatment.
	
	The more global picture is such that the noise is close to the
	Poissonian level, whenever the current is relatively large.  Thus
	possible applications and measurements of a ratchet current induced by
	tunnel oscillations, should not be hindered by current fluctuations.

	\begin{acknowledgments}
		We thank F. Dom{\'i}nguez for helpful discussions.
		This work has been supported by the
		Spanish Ministry of Economy and Competitiveness via a
		FPU scholarship (R.H.) and through Grant No.\ MAT2011-24331.
	\end{acknowledgments}
			
	\appendix
	\section{\label{sec.:Liouvillian}Liouvillian and jump operators}
	For a system-bath Hamiltonian, one can derive
	for the reduced system density operator $\rho$ the Bloch-Redfield or
	Born-Markov master equation~\cite{Blum1996a, Carmichael2002a, Breuer2003a}
	\begin{align}
		\begin{split}
\dot\rho(t) =
	&-i\com{\hat H_S}{\rho(t)} \\
	&-\int_0^\infty ds \tr_B \com{\hat H_V}{\com{\tilde H_V(-s)}{\rho(t)R_0}}
\\ \equiv{}  & \mathcal L_0\rho(t) ,
		\end{split}
\label{app:ME}
	\end{align}
	where $R_0$ is the reference density operator of the environment,
	while $\hat H_V$ describes the system-bath coupling.  It has to be weak,
	such that coherent system dynamics dominates.
	Augmenting the environment density operator by a counting variable for
	the lead electrons according to $\rho(t)R_0 \to \rho(t)e^{i\bm\chi
	(\hat{\bm  N}-\bm N_0)}R_0$, yields the $\bm{\chi}$-dependent density operator
	$\rho(\bm{\chi},t)$ whose trace is the moment-generating function
	introduced in Sec.~\ref{sec.:eff_masterequation}.\cite{KaiserAP2007a}  Furthermore,
	one obtains by inserting the same ansatz into Eq.~\eqref{app:ME} the
	master equation \eqref{eq.:DQS_masterequation} upon which all our
	results are based.
	
	For the evaluation of the time integral in Eq.~\eqref{app:ME}, it is
	convenient to work in the eigenbasis of the system Hamiltonian,
	defined by $E_k\ket{m} = \hat H_S\ket{m}$.  Then one obtains for
	the density matrix elements the equation of motion
	$
	\dot\rho_{mn}(\bm{\chi},t) = \sum_{kl} [\mathcal L(\bm{\chi})]_{mn,kl} \;\rho_{kl}(\bm{\chi},t)
	$,
	where the decomposed Liouvillian
	\begin{align}
[\mathcal L(\bm{\chi})]_{mn,kl} = [\mathcal L_0]_{mn,kl}
+\sum_{\alpha,s=\pm}(e^{s i\chi_\alpha}-1)[\mathcal J^s_\alpha]_{mn,kl}
	\end{align} 
	consists of the contributions
	\begin{align}
[\mathcal L_0]_{mn,kl} 
	&= 	-i\delta_{km}\delta_{ln}(E_m - E_n)
		+\sum_{\alpha,s=\pm}[\mathcal J^s_\alpha]_{mn,kl}
		\nonumber\\
	&\hspace{2ex} -\frac{1}{2}\delta_{ln} 
		\sum_{\alpha,p}\gamma_{\alpha,pk} [\hat d_\alpha]_{mp} [\hat d_\alpha^\dag]_{pk} \nonumber\\
	&\hspace{2ex} -\frac{1}{2}\delta_{km}  
		\sum_{\alpha,p}\gamma_{\alpha,pl} [\hat d_\alpha]_{lp} [\hat d_\alpha^\dag]_{pn} \nonumber\\  
	&\hspace{2ex} -\frac{1}{2}\delta_{ln}  
		\sum_{\alpha,p}\bar\gamma_{\alpha,kp} [\hat d_\alpha^\dag]_{mp} [\hat d_\alpha]_{pk} \nonumber\\
	&\hspace{2ex} -\frac{1}{2}\delta_{km}  
		\sum_{\alpha,p}\bar\gamma_{\alpha,lp} [\hat d_\alpha^\dag]_{lp} [\hat d_\alpha]_{pn},\\
[\mathcal J^-_\alpha]_{mn,kl} 
	&= \frac{1}{2}(\gamma_{\alpha,mk}+\gamma_{\alpha,nl})
		[\hat d_\alpha^\dag]_{mk} [\hat d_\alpha]_{ln},\\
[\mathcal J^+_\alpha]_{mn,kl}
	&= \frac{1}{2}(\bar\gamma_{\alpha,km}+\bar\gamma_{\alpha,ln})
		[\hat d_\alpha]_{mk} [\hat d_\alpha^\dag]_{ln}.
	\end{align} 
	The effective rates
	\begin{align}
\gamma_{\alpha,mn} &= \Gamma_\alpha f_\alpha(E_m-E_n),\\
\bar \gamma_{\alpha,mn} &= \Gamma_\alpha [1- f_\alpha(E_m-E_n)],
	\end{align}
	depend on the Fermi functions of the leads, $f_\alpha(\omega) \equiv f(\omega - \mu_\alpha) =
	\{\exp[(\omega - \mu_\alpha)/k_BT]+1\}^{-1}$, while
	\begin{align}
\Gamma_\alpha &=
	2\pi\sum_{k}\abs{V_{k\alpha}}^2\delta(\omega-\varepsilon_{k\alpha})
	\end{align}
	denotes the spectral densities of the dot-lead couplings, which within a
	wide-band limit are assumed energy independent.

	\section{\label{sec.:zero_bias}Spectral decomposition of the ratchet Liouvillian}
	In the energy basis 
	$\{\ketbra{0}{0}, \ketbra{e}{e}, \ketbra{g}{g},
	\ketbra{g}{e}, \ketbra{e}{g}\}$
	the ratchet Liouvillian for vanishing counting variable $\chi_2\to0$ reads
	\begin{align}
	\mathcal L_{\textrm{ra}}
			= \begin{pmatrix}
		-\Gamma_\textrm{ra} & \Gamma_\textrm{ra} & 0 & 0 & 0\\
		0 & -\Gamma_\textrm{ra} & 0 & 0 & 0\\
		\Gamma_\textrm{ra} & 0 & 0 & 0 & 0\\
		0 & 0 & 0 & -\frac{\Gamma_\textrm{ra}}{2} +i\delta & 0\\
		0 & 0 & 0  & 0 & -\frac{\Gamma_\textrm{ra}}{2} -i\delta\\
			\end{pmatrix}.
	\end{align}
	Within the perturbative treatment of Sec.~\ref{sec.:eff_masterequation},
	we need to compute functions $f(\mathcal{L}_\text{ra})$ of this matrix,
	such as the propagator $\exp(\mathcal{L}_\text{ra}t)$ or the resolvent
	$(z-\mathcal{L}_\text{ra})^{-1}$, which is usually achieved by
	spectral decomposition of the Liouvillian.  Here however this is hindered by the fact
	that $\mathcal{L}_\text{ra}$ is defective, i.e., it does not possess
	a complete set of eigenvectors.  The problem arises from the upper block
	\begin{align}
			L \equiv \begin{pmatrix}
		-\Gamma_\textrm{ra} & \Gamma_\textrm{ra} & 0\\
		0 & -\Gamma_\textrm{ra} & 0\\
		\Gamma_\textrm{ra} & 0 & 0\\
			\end{pmatrix} ,
	\end{align}
	which we transform via
	\begin{align}
	S = 
			\begin{pmatrix}
		0 & -1 & 0\\
		0 & 0 & -{1}/{\Gamma_\textrm{ra}}\\
		1 & 1 & {1}/{\Gamma_\textrm{ra}}
			\end{pmatrix} ,
	\end{align}
	to the Jordan canonical form\cite{GolubSIAMReview1976a,MolerSIAMReview1978a}%
	\begin{equation}
		J = S^{-1} L S
		=
		\begin{pmatrix}
			0 & 0 & 0\\
			0 & -\Gamma_\textrm{ra} & 1\\
			0 & 0  & -\Gamma_\textrm{ra}
		\end{pmatrix}		.
	\end{equation}
	Its eigenvalues obviously are $0$ and the twofold degenerate
	$-\Gamma_\text{ra}$, and one immediately finds two vectors that
	obey the eigenvalue equation, namely
	\begin{align}
	L\ket{0} ={} & 0\ket{0},\\
	L\ket{1} ={} & -\Gamma_\textrm{ra}\ket{1}.
	\end{align}
	
	A generalized eigenbasis can be found by including a third vector $|2\rangle$
	that fulfills \cite{GolubSIAMReview1976a,MolerSIAMReview1978a}
	\begin{equation}
		L\ket{2} = -\Gamma_\textrm{ra}\ket{2} + \ket{1},
	\end{equation}
	i.e., one adds the eigenvector of the degenerate subspace.
	By repeated multiplication with $L$ follows
	$L^k\ket{2}=(-\Gamma_\textrm{ra})^k\ket{2} +
	k(-\Gamma_\textrm{ra})^{k-1}\ket{1}$, which implies
	\begin{align}
	f(L)\ket{2} &= f(\lambda_2)\ket{2} +  f'(\lambda_2)\ket{1},
	\end{align}
	where both the function of a matrix and its derivative are
	defined as the corresponding Taylor series.  This relation
	together with the usual $f(L)\ket{k} = f(\lambda_k)\ket{k}$
	for $k=0,1$, allows us to evaluate any $f(L)$.
	In particular,
	we find the propagator
	\begin{align}
			e^{Lt} = &\begin{pmatrix}
		e^{-\Gamma_\textrm{ra}t} & \Gamma_\textrm{ra}t e^{-\Gamma_\textrm{ra}t}  & 0\\
		0 & e^{-\Gamma_\textrm{ra}t} & 0\\
		1-e^{-\Gamma_\textrm{ra}t}  & 1-(1+\Gamma_\textrm{ra}t)e^{-\Gamma_\textrm{ra}t}  & 1\\
			\end{pmatrix},
	\end{align}
	and the resolvent
	\begin{align}
			(z-L)^{-1} = &\begin{pmatrix}
		\frac{1}{z+\Gamma_\textrm{ra}} & \frac{\Gamma_\textrm{ra}}{(z+\Gamma_\textrm{ra})^2}  & 0\\
		0 & \frac{1}{z+\Gamma_\textrm{ra}}  & 0\\
		\frac{\Gamma_\textrm{ra}}{z(z+\Gamma_\textrm{ra})}  & \frac{\Gamma_\textrm{ra}^2}{z(z+\Gamma_\textrm{ra})^2}  & \frac{1}{z}\\
			\end{pmatrix}.
	\end{align}
	
	A poor man's approach to this procedure \cite{MolerSIAMReview1978a} is
	to introduce a small perturbation that lifts the degeneracy of $L$.
	After evaluating $f(L)$, one considers the limit of vanishing
	perturbation.
	
	\section{\label{sec.:occupation_cumulant}Stationary state of the drive circuit}
	The effective ratchet Liouvillian derived in
	Sec.~\ref{sec.:eff_masterequation} results from a perturbation theory
	with the stationary states representing the zeroth order.
	They are determined by the master equation\cite{GurvitzPRB1996a}
	\begin{align}
\label{app:driveME}
	\dot\rho_\textrm{dr} 
	&= -i\com{\hat H_{S,\textrm{dr}}}{\rho_\textrm{dr}} 
		+\mathcal D(\hat d_3^\dag)\rho_\textrm{dr}
		+\mathcal D(\hat d_4)\rho_\textrm{dr},
	\end{align}
	with the system Hamiltonian of the drive
	\begin{equation}
		\hat H_{S,\textrm{dr}} 
		=-\big(T_\text{dr} \hat d_3^\dag\hat d_4 + T_\text{dr}^* \hat d_4^\dag\hat d_3\big) .
	\end{equation}
	The last two terms in the master equation describe dot-lead tunneling,
	which in the limit of a large voltage bias obeys the Lindblad form
	\begin{equation}
		\mathcal D(\hat d_4)\rho_\textrm{dr}
		= \Gamma_{\textrm{dr}}\;\hat d_4 \rho_\textrm{dr} \hat d_4^\dag
		-\frac{\Gamma_{\textrm{dr}}}{2} \acom{\hat d_4^\dag\hat d_4}{\rho_\textrm{dr}}.
	\end{equation}
	For the drive circuit, the stationary state can be obtained
	conveniently after a decomposition of the Liouvillian into the
	corresponding Fock basis, by which we find
	\begin{align}
	\rho_\textrm{dr}^\textrm{stat} = {}&\frac{1}{\Gamma_{\textrm{dr}}^2+12\abs{T_\textrm{dr}}^2}\nonumber\\
			&\times\begin{pmatrix}
		4\abs{T_\textrm{dr}}^2 & 0 & 0\\
		0 & \Gamma_{\textrm{dr}}^2+4\abs{T_\textrm{dr}}^2 & -2i\Gamma_{\textrm{dr}}T_\textrm{dr}^*\\
		0 & 2i\Gamma_{\textrm{dr}}T_\textrm{dr} & 4\abs{T_\textrm{dr}}^2
			\end{pmatrix}.
	\end{align}
	
	The auto correlation function of the population imbalance $\Delta \hat
	n_\textrm{dr} = \hat n_4 - \hat n_3$ in Laplace space is defined as
	\begin{align}
\label{Cz}
C(z)
& = \int_{0}^{\infty}dt\;e^{-zt}\big[\ev{\Delta\tilde n_{\textrm{dr}}(t)\Delta\tilde n_\textrm{dr}(0)} 
	- \ev{\Delta\hat n_{\textrm{dr}}}^2\big]\nonumber\\
&=\ev{\Delta\tilde n_{\textrm{dr}}(z)\Delta\hat n_{\textrm{dr}}}
	- \frac{1}{z}\ev{\Delta\hat n_{\textrm{dr}}}^2,
	\end{align}
	with the stationary occupation
	\begin{align}
\ev{\Delta\hat n_{\textrm{dr}}}
&= -\frac{\Gamma_{\textrm{dr}}^2}{\Gamma_{\textrm{dr}}^2+12\abs{T_\textrm{dr}}^2}
	\end{align}	
	and the corresponding correlation function
	\begin{align}
&\ev{\Delta\tilde n_{\textrm{dr}}(z)\Delta\hat n_{\textrm{dr}}}
= \frac{1}{z}\frac{2z + \Gamma_{\textrm{dr}}}{\Gamma_{\textrm{dr}}^2+12\abs{T_\textrm{dr}}^2}\nonumber\\
&\hspace{1em}\times\frac{
		\Gamma_{\textrm{dr}}^2(z+\Gamma_{\textrm{dr}})^2 +4z(2z+3\Gamma_{\textrm{dr}})\abs{T_\textrm{dr}}^2
	}{
		(z+\Gamma_{\textrm{dr}})^2(2z+\Gamma_{\textrm{dr}}) +4(2z+3\Gamma_{\textrm{dr}})\abs{T_\textrm{dr}}^2
	}.
	\end{align}

\end{document}